\documentclass[twocolumn]{autart}

\usepackage{amsfonts,amssymb,amstext,verbatim,amsmath,color}
\usepackage[T1]{fontenc}
\usepackage{graphicx}

\usepackage[font=small,labelfont=bf]{caption}

\newtheorem{theorem}{\textbf{Theorem}}
\newtheorem{lemma}[theorem]{\textbf{Lemma}}

\newtheorem{remark}{\textbf{Remark}}

\newtheorem{algo}{\textbf{Algorithm}}

\newcommand{\ba}{\begin{array}}
\newcommand{\ea}{\end{array}}

\newcommand{\nat}{\mathbb{N}}

\newcommand{\rea}{\mathbb{R}}

\begin{document}
\begin{frontmatter}
\title{Decentralized Estimation of Laplacian Eigenvalues \\ in Multi-Agent Systems}

\thanks{The research leading to these results has received funding from the
European Union Seventh Framework Programme [FP7/2007-2013] under
grant agreement n 257462 HYCON2 Network of excellence.
}

\thanks{The research leading to these results has received funding from the Italian grant FIRB Futuro in Ricerca, project NECTAR, code
RBFR08QWUV.}

\author[Cagliari]{Mauro Franceschelli}\ead{mauro.franceschelli@diee.unica.it},
\author[Rome]{Andrea Gasparri}\ead{gasparri@dia.uniroma3.it},
\author[Cagliari]{Alessandro Giua}\ead{giua@diee.unica.it},
\author[Cagliari]{Carla Seatzu}\ead{seatzu@diee.unica.it}.

\address[Cagliari]{Department of Electrical and Electronic Engineering, University of Cagliari, Piazza D'Armi, 09123 Cagliari, Italy}  % Please supply
\address[Rome]{Department of Computer Science and Automation, University of ``Roma Tre'', Via Vasca Navale 79, 00146 Roma, Italy}             % full addresses

\begin{keyword}                           % Five to ten keywords,
Laplacian spectrum, decentralized estimation, multi-agent systems.               % chosen from the IFAC
\end{keyword}

\maketitle

\begin{abstract}
In this paper we present a decentralized algorithm to estimate the eigenvalues of the Laplacian matrix that encodes the network topology of a
multi-agent system. \text{{We consider network topologies modeled by undirected graphs}}. The basic idea is to provide a local interaction rule among agents so that their state trajectory is a linear combination of sinusoids oscillating only at frequencies function of the eigenvalues of the  Laplacian matrix. In this way, the problem of decentralized \text{{estimation of the eigenvalues}} is mapped into a standard signal processing problem in which the unknowns are the finite number of frequencies at which the signal oscillates.
\end{abstract}

\end{frontmatter}
%\begin{IEEEkeywords}
%Laplacian Spectrum, Laplacian Eigenvectors, Multi-Agent System.
%\end{IEEEkeywords}

%\IEEEpeerreviewmaketitle

\section{Introduction}
%Nowadays {the research community is investing more and more effort in} designing coordination and estimation algorithms for networked multi-agent systems \cite{BHOT05,bullo06,Jad03,OSaber04}.  %Such systems represent an ideal abstraction of actual networks of mobile robots that are envisioned to  perform the most various kind of tasks in the future.
%{Networks of agents are described by graphs, where nodes represent agents and edges represent couplings between them \cite{me04}.  %e.g., proximity graphs \cite{bullo06}.
%The emergent behavior of such a system depends both on the interactions between agents and on the network topology.
%
%Algebraic graph theory \cite{godsil} provides powerful tools to characterize interesting properties of the network topology by encoding a graph in a matrix.
%%A {significant} effort has been made to characterize the topological properties of {graphs}.
%As an example, the knowledge of the spectrum of the Laplacian matrix associated to a graph can be used to estimate topological properties of an undirected graph \cite{Mohar:1991}. Furthermore the spectrum of the Laplacian provides additional information on the dynamical properties of a networked control system. As an example, as stated in \cite{OSaber04}, the second smallest eigenvalue also called algebraic connectivity is a fundamental parameter to estimate the worst case convergence rate of consensus algorithms and, more in general, of multi-agent systems with local interactions described by the Laplacian matrix such as leader-follower networks \cite{ren2007consensus}.}

Nowadays {the research community is investing more and more effort in} designing coordination and estimation algorithms for networked multi-agent systems \cite{BHOT05,bullo06,Jad03,OSaber04}.   {The network topology of a multi-agent system can be effectively described by means of a graph, where nodes represent agents and edges represent couplings between them \cite{me04}.  The emergent behavior of such a system depends both on the interactions between agents and on the network topology.

Algebraic graph theory \cite{godsil} provides powerful tools to analyze a graph. As an example, the knowledge of the spectrum of the Laplacian matrix associated to a graph can be used to estimate topological properties of an undirected graph,  e.g., algebraic connectivity, average degree, diameter, spectral gap, connectivity measures \cite{Mohar:1991}. In the context of multi-agent systems, this knowledge may also provide powerful insights into the dynamics of a networked control system. As an example, as stated in \cite{OSaber04}, the algebraic connectivity, i.e., the second smallest eigenvalue, is a fundamental parameter to estimate the worst case convergence rate of consensus algorithms and, more in general, of multi-agent systems with local interactions described by the Laplacian matrix such as leader-follower networks \cite{ren2007consensus}.}

%As an example, in the context of multi-agent system, the knowledge of the spectrum of the Laplacian matrix associated to a graph can be used to estimate topological properties of an undirected graph \cite{Mohar:1991}. Furthermore the spectrum of the Laplacian provides additional information on the dynamical properties of a networked control system. As an example, as stated in \cite{OSaber04}, the second smallest eigenvalue also called algebraic connectivity is a fundamental parameter to estimate the worst case convergence rate of consensus algorithms and, more in general, of multi-agent systems with local interactions described by the Laplacian matrix such as leader-follower networks \cite{ren2007consensus}.}

 %For a comprehensive overview of the topological {properties retrieved} from the spectrum of the Laplacian matrix we refer to \cite{Mohar:1991} and references therein.}

{Unfortunately, the spectrum of the Laplacian matrix is not readily computable in a distributed setting where the network topology is unknown.}
In order to overcome this limitation, we have designed a local interaction rule so that the resulting dynamical system oscillates only at frequencies corresponding to the eigenvalues of the Laplacian matrix that encodes the network topology. In this way,  {the problem of estimating the eigenvalues} is mapped into a signal processing problem solvable independently by each agent in a decentralized fashion, using tools from signal processing or system identification theory.

 Compared to the state of the art, discussed in detail in next section, our approach allows to estimate the full spectrum of the {symmetric} Laplacian matrix without the need to estimate all the corresponding eigenvectors.  Moreover our approach provides an {approximate estimation} of the eigenvalues in finite time.

The contributions of this paper are the following:
\begin{itemize}
\item We propose a novel local interaction rule to make the network state oscillate at frequencies corresponding to the eigenvalues of the Laplacian matrix, thus mapping the decentralized eigenvalue estimation problem into a standard signal processing problem.
\item We extend \cite{Franceschelli:2009} by characterizing analytically the amplitude and phase of the oscillations as function of the eigenvectors of the Laplacian and the initial conditions.
\item We propose an improvement with respect to \cite{Franceschelli:2009} so that no component at null frequency exists in the evolution of the agents' state.  The removal of the DC component allows the straightforward application of frequency estimation algorithms such as the one in \cite{HybridAdaptiveFreq:11}.
%\item We integrate the method in \cite{HybridAdaptiveFreq:11} for estimating the frequencies at which a signal composed only by sinusoids oscillates into our approach.
\end{itemize}
\subsection*{Related works}
In \cite{zavlanos2011spectral} Zavlanos \emph{et al.} {investigated the problem of how to coordinate a network of mobile robots with position-dependent topology so that the corresponding adjacency matrix has a given set of eigenvalues. This}  approach is based on artificial potentials, function of the inter-agent distances, that allow a gradient descent algorithm to make the network converge to a topology whose eigenvalues are the desired ones. In this preliminary paper the authors consider the spectral moments related to the spectrum of the adjacency matrix to be known.

 In \cite{franceschelli2010observability} Franceschelli \emph{et al.} presented a necessary and sufficient condition {to verify observability and controllability with of a leader-follower network with unknown topology of mobile vehicles is proposed based on the algorithm} in \cite{Franceschelli:2009} and its extension in this paper.

 In \cite{Gordon:2008} Yang \emph{et al.} proposed a technique for the estimation of the second smallest eigenvalue  of a weighted Laplacian matrix based on the power iteration algorithm by the estimation of the corresponding eigenvector. In addition, the authors discuss a decentralized control algorithm to maximize the algebraic connectivity. The idea is to let agents move so that links are added or weights changed as two agents come closer.% Indeed, this requires agents to exchange their positions; thus a common reference frame for the agents must be available to run the algorithm.
% In \cite{Jadbabaie:2006} the authors propose a decentralized super-gradient algorithm for the maximization of the algebraic connectivity, while the control action is based on virtual potential fields. The key idea is to first iteratively compute the optimum Laplacian matrix and then let the agents move towards the corresponding topology. Also in this approach, the maximization process requires the estimation of the  eigenvector related to the second smallest eigenvalue of the
%Laplacian graph, which is carried out by means of the Decentralized Orthogonal Iteration Algorithm. Nevertheless, compared to the previous approach, a smaller amount of data, i.e., the eigenvector components related to the neighborhood,  has to be exchanged among agents.

 In \cite{Sahai201215} Sahai \emph{et al.} presented an approach building on the idea of \cite{Franceschelli:2009} for the application of clustering. The authors propose a local interaction rule formally equivalent to the wave equation discretized in time and space and show that the ``wave equation method'', can be used to cluster a graph by estimating the sign of the coefficients of the discrete Fourier transform corresponding to the second smallest eigenvalue. Furthermore, their approach is superior with respect to the convergence speed to the state of the art of eigenvector based clustering algorithms.

 {Finally, in \cite{Kempe200870} Kempe \emph{et al.} are interested in computing in a decentralized
way an approximation of the first $k$ eigenvectors of a symmetric matrix that encodes the network topology. Their algorithm takes inspiration from the orthogonal iteration algorithm and assumes that the network topology is unknown to the nodes. This algorithm can be adapted to our objective, i.e., the distributed estimation of the eigenvalues of the Laplacian matrix, by introducing a distributed technique for the computation of the Rayleigh quotient. }

\section{Online Spectrum Estimation} \label{sec3}
Let us consider {the interactions of a network of agents described} by an {undirected} graph $\mathcal{G}=\{V,\mathcal{E}\}$, where $V=\{1,\ldots,n\}$ is the set of agents and $\mathcal{ E}\subseteq V \times V$ is the set of edges: an edge $e_{i,j}$ exists between agents $i$ and $j$ if agent $i$ interacts with agent $j$.

 Let $\mathcal{N}_{i}$ define the neighborhood of agent $i$, namely the set of indices of the agents
connected by an edge with agent $i$. In particular, $|\mathcal{ N}_{i}|=\Delta_i$
where $\Delta_{i}$ is called degree of agent $i$. Let $\mathcal{ L}$ be the Laplacian matrix of graph $\mathcal{ G}$, it is a $n \times n$  matrix the {elements of which} are $l_{ij}=\Delta_{i}$ if $i=j$, $l_{ij}=-1$  if  $j \in \mathcal{N}_{i}$ and $0$ otherwise. The Laplacian matrix $\mathcal{L}$ of {an undirected graph} is symmetric by construction and thus all its eigenvalues are real.  Furthermore, {for a connected graph}, it has one null structural eigenvalue with corresponding eigenvector equal to the vector of ones $\textbf{1}_{n}$ of appropriate dimensions, thus $\mathcal{L}\,\textbf{1}_{n}=\textbf{0}_{n}$. %The number of null eigenvalues corresponds to the number of connected components of $\mathcal{ G}$ and $Rank(\mathcal{L})=n-c,$ where $n$ is the number of nodes and $c$ is the number of connected components of $\mathcal{ G}$ \cite{OSaber04}.
In addition, according to the Gershgorin disc theorem, {a symmetric Laplacian has all its eigenvalues located within} $[0, 2\, \Delta_{max}]$, where $\Delta_{max}$ is the maximum degree between the agents in the graph.

 We now present a decentralized algorithm to estimate the eigenvalues of the Laplacian matrix.
The algorithm requires each agent $i$ to store two variables  $x_i,z_i\in \rea$ and apply a local update rule upon receiving the values of the equivalent variables from its neighbors.
\begin{algo}[Online Spectrum Estimation]\hfill \label{spectrumest}
\begin{enumerate}
\item Each agent sets $t=0$ and chooses an initial condition uniformly at random $x_i(0),z_i(0)\in \left \{ -1,1\right \}$.
\item Each agent \emph{simulates} the following local interaction rule with its neighbors $\mathcal{ N}_i(t)$
\begin{equation} \label{updaterule}
\left\{ \begin{array} {l}
\dot{x}_i(t)  = z_i(t)+  \sum_{j\in \mathcal{ N}_i} \left(z_i(t) - z_j(t)\right), \\
\dot{z}_i(t)  = -x_i(t) - \sum_{j\in \mathcal{ N}_i} \left(x_i(t) - x_j(t)\right). \\
\end{array}
\right.
\end{equation}
\item In a time window of length $T$, agent $i$ estimates the frequencies of the sinusoids of which signal $x_i(t)$ is composed.
\item The values of the frequencies estimated correspond to the eigenvalues of the Laplacian matrix $\mathcal{L}$ shifted by $1$ and are given as output. \hfill $\blacksquare$
\end{enumerate}
\end{algo}

Note that $Step \ 3$ can be solved by several methods of signal processing or system identification. In particular, the {required} value of $T$ depends on the chosen algorithm. In this paper, as discussed in Section~\ref{sec6}, we exploit the method presented in \cite{HybridAdaptiveFreq:11} to implement Algorithm~\ref{spectrumest}.
% that requires $T$ to be greater than the period of the smallest frequency that in our case is by Lemma~\ref{lemma:eigenvalues} always equal to $1$.
% The proposed algorithm requires that each agent performs the state updating rule~\eqref{updaterule}.
%

The behavior of the network  when all the agents update their state according to eq.~\eqref{updaterule} {can be described as follows}
\begin{equation} \label{linsys}
\left[\begin{array} {c}
\dot{x}(t) \\
\dot{z}(t)
\end{array}\right] =
\mathcal{ A} \cdot \left[
\begin{array} {c}
x(t) \\
z(t)
\end{array}\right], \quad
\mathcal{ A} =\left[\begin{array} {cc}
\textbf{0}_{n \times n} & I+\mathcal{ L}\\
-I-\mathcal{ L} & \textbf{0}_{n \times n}
\end{array} \right].
%\label{linsys}
\end{equation}
{where} $I$ is the $n \times n$ identity matrix and $\textbf{0}_{n \times n}$ is the null $n \times n$ matrix.
Note that for any network topology $\mathcal{A}$ is skew symmetric, i.e., $\mathcal{ A}^T=-\mathcal{ A}$.
In the following theorem, we prove that the eigenvalues of $\mathcal{ A}$ can be analytically derived from the eigenvalues of the
Laplacian matrix $\mathcal{ L}$ and they are all structurally purely imaginary. %It is relevant to point out that since interactions depend only on the network topology, no uncertainty is present in the coefficients of matrix ${\mathcal A}$.
\begin{lemma} \label{lemma:eigenvalues}
Let $\mathcal{ G}$ be an undirected graph with Laplacian $\mathcal{ L}$. Let
matrix $\mathcal{ A}$ be defined as in eq.~\eqref{linsys}. To any
eigenvalue $\lambda_{\mathcal{L}}$ of $\mathcal{ L}$ it corresponds a
couple of complex and conjugate eigenvalues $\lambda_\mathcal{ A},\bar \lambda_\mathcal{ A}$ of $\mathcal{ A}$, that is:
$$\lambda_\mathcal{ A}= j(1 + \lambda_{\mathcal{L}}), \quad \bar \lambda_\mathcal{ A}=-j( 1 + \lambda_{\mathcal{L}}),$$
while the corresponding eigenvectors $v_{\lambda_{\mathcal{A}}}$  are function of the eigenvectors $v_{\lambda_{\mathcal{ L}}}$ of $\mathcal{ L}$
$$
v_{\lambda_{\mathcal{ A}}} = \big [ v_{\lambda_{\mathcal{ L}}}^{T} \quad jv_{\lambda_{\mathcal{ L}}}^{T}  \big ]^T,
\qquad
\bar v_{\bar \lambda_{\mathcal{ A}}}  = \big [ v_{\lambda_{\mathcal{ L}}}^{T} \quad -j v_{\lambda_{\mathcal{ L}}}^{T} \big ]^T.
$$

 \rm \emph{Proof:} See Appendix \ref{appendix:teoeigenvectors}. \hfill$\square$
\end{lemma}
By Lemma~\ref{lemma:eigenvalues} it follows that the state of each agent has an oscillatory trajectory. Furthermore, as detailed by Theorem~\ref{teo5}, this trajectory is a linear combination of sinusoids oscillating only at frequencies function of the eigenvalues of the matrix Laplacian.
In the following we assume the Laplacian to have  $m$ \emph{distinct} eigenvalues  labeled as follows: $ 0=\lambda_1 < \lambda_2 < \cdots < \lambda_m$.
\begin{theorem} \label{teo5}
Let us consider a system described by eq.~\eqref{linsys} relative to
a network whose graph $\mathcal{ G}$ is connected.   Let $x(0)=x_0$ and $z(0)=z_0$ be the state
initial conditions. Let $\delta(\cdot)$ be the Dirac's delta function. Let $\lambda_{j}$ be an eigenvalue of the Laplacian matrix $\mathcal{L}$ of graph $\mathcal{ G}$ with algebraic multiplicity $\nu_j$ and let $m$ be the number of distinct eigenvalues. Let $v_1$ be the unitary norm eigenvector corresponding to $\lambda_1=0$, and $v_{j}^{(k)}$, \mbox{$k=1,\ldots, \nu_j$,} be the
$\nu_j$ unitary norm eigenvectors associated to $\lambda_j>0$.
The module of the Fourier transform of the $i$-th state components $x_i(t)$ and $z_i(t)$, \mbox{$i=1,\ldots, n$,} can be written as
$$|\mathcal{ F}[x_i(t)]|=|X_i(f)|= \sum_{j=1}^m   \frac{a_{j,i}}{2}   \, \delta \left (f \pm \frac{1 + \lambda_j }{2\pi} \right ),$$
$$|\mathcal{ F}[z_i(t)]|=|Z_i(f)|= \sum_{j=1}^m  \frac{b_{j,i}}{2}   \,\delta \left ( f \pm \frac{1 + \lambda_j }{2\pi} \right ),$$
{where $f$ is the frequency domain variable.}
In addition, the coefficients $a_{j,i}$ and $b_{j,i}$ are given by

 -- For $\lambda_{ 1} = 0$ %{ ($\nu_{ 1} = 1$ since the graph is connected by assumption)}:
\begin{equation}\label{coefficients}
\begin{array}{llllll}
a_{ 1,i} &=& v_{1}(i) \,
v_{1}^Tx(0)=\dfrac{\textbf{1}_{n}^T \, x(0)}{n}, \\
b_{ 1,i}  &=&  v_{1}(i) \,
v_{1}^Tz(0)=\dfrac{\textbf{1}_{n}^T \, z(0)}{n}.
\end{array}
\end{equation}
 -- For $\lambda_j  > 0$ %and $\nu_{j} \geq 1$ :
\begin{equation}\label{coefficients2}
{a_{j,i}= b_{j,i}=  \sqrt{ \left [ \sum_{k=1}^{\nu_{j}} \Big (v^{(k)}_{j}(i)    {v^{(k)}_{j}}^Tx(0) \Big ) \right ]^{2} + \atop \left [\sum_{k=1}^{\nu_{j}} \Big (v^{(k)}_{j}(i) {v^{(k)}_{j}}^Tz(0) \Big ) \right ]^{2} }.}%\\~\\
\end{equation}

\rm \emph{Proof:} See Appendix \ref{appendix:teo5}.  \hfill $\square$
\end{theorem}
The above theorem states the key result of this paper. In fact, it implies that each agent can independently  solve
the {problem estimating the eigenvalues} by estimating the frequencies at which its own state variable $x_i(t)$ oscillates.
%
% This approach allows in principle to detect in a decentralized way \textit{changes} in the network topology, such as network
%disconnections, changing in the formation of the agents, loss of crucial links that decrease the network algebraic connectivity and so on.
\begin{remark}
Few important remarks are now in order:
\begin{itemize}
\item[-] The value of $x_i(t)$ can be seen as the output of the $i$-th agent.
{If the system} is not observable from the output $x_i(t)$ then some coefficients $a_j$ are null and thus the corresponding mode
cannot be detected by agent $i$. %This implies that the {corresponding eigenvalue $\lambda_j$ can not be} estimated by agent $i$. The observability properties of system~\eqref{linsys} are addressed in the following Theorem~\ref{teo:obs}.
\item [-] For each agent, the amplitude of the sinusoid oscillating at $\omega=\lambda_1=1$ corresponds to the instantaneous average of the state variables. \hfill $\blacksquare$
\end{itemize}
\end{remark}
The observability and controllability of a system, the dynamics of which are described by the Laplacian {matrix, have been studied} in \cite{MMA07,me07} from a graph theoretical point of view. {In the following theorem} we link the ability to estimate all the eigenvalues of the Laplacian to its observability property.
\begin{theorem} \label{teo:obs}
Let $\mathcal{ A}$ be the matrix describing the group dynamics as in \eqref{linsys}. Let $C$ be a $k\times n$ output matrix, with $k\in \nat$.
 Let
$$\mathcal{ A} = \left[ \begin{array}{cc}
                    \textbf{0}_{n \times n} & I+\mathcal{ L} \\
                    -I-\mathcal{ L} & \textbf{0}_{n \times n}
                  \end{array} \right] \quad \text{and} \quad \hat C = \left[ \begin{array}{cc}
                    C & \textbf{0}_{k \times n} \\
                    \textbf{0}_{k \times n} & C
                  \end{array} \right].
$$
 Let $\mathcal{O}_{\mathcal{A}}=\mathcal{O}(\mathcal{A}, \hat{C})$ and $\mathcal{O}_{\mathcal{L}}=\mathcal{O}(\mathcal{L}, C)$ be the {observability matrices} built with the corresponding matrices. Then: $$Rank\left(\mathcal{O}_{\mathcal{A}}\right)=2 Rank\left(\mathcal{O}_{\mathcal{L}}\right).$$

 \rm \emph{Proof:} See Appendix~\ref{appendix:teo:obs}. \hfill $\square$
\end{theorem}
%
%Moreover, even if each agent cannot observe all the eigenvalues for a given topology by exploiting the proposed method, proper algorithms could be developed to spread this information over the network.
%
 We now state the main result of this paper that proves the correctness of Algorithm~\ref{spectrumest}.
\begin{theorem}
Consider a connected network $\mathcal{G}$ of $n$ agents that executes Algorithm~\ref{spectrumest}. Let the initial conditions of system~\eqref{linsys} be not orthogonal to {any eigenvector} of matrix $\mathcal{L}$.  Let $C=\left[0 \ldots, 1,\ldots, 0\right]$ be zero everywhere except for the $i$-th unitary element, with $i\in V$. If $\mathcal{O}\left(\mathcal{L},C\right)$ is full rank, then agent $i$ can estimate all the eigenvalues of the Laplacian matrix.

 \rm \emph{Proof:} Due to Lemma~\ref{lemma:eigenvalues} all the eigenvalues of system~\eqref{linsys} are purely imaginary and correspond to the eigenvalues of the Laplacian matrix shifted by one. Furthermore, if the initial conditions are not orthogonal to all the eigenvectors of $\mathcal{L}$ and $\mathcal{O}\left(\mathcal{L},C\right)$ is full rank as discussed in Theorem~\ref{teo:obs}, then all the sinusoids corresponding to the system modes have coefficients strictly greater than zero. Thus by applying a frequency estimation algorithm to the signal $x_i(t)$, for instance the one in \cite{HybridAdaptiveFreq:11}, agent $i$ can estimate the full spectrum of the Laplacian matrix by only observing its own state evolution. \hfill $\square$
\end{theorem}

It is relevant to point out that even if the system is not
observable from a single agent perspective, it will always be
observable if matrix $C$ is the identity matrix, i.e., if we
consider all the information that agents locally retrieve.

\section{Numerical implementation of the approach} \label{sec6}
The system in eq.~\eqref{linsys} is a marginally stable linear system since all its eigenvalues lie on the imaginary axis. The {stability of a system with} eigenvalues exactly on the imaginary axis is not considered to be robust because even the slightest parameter uncertainties may render the system unstable. In our case there is no parameter uncertainty because system \eqref{linsys} is based on the Laplacian matrix {the elements of which} depend only on the {existence} of links between the agents. {Thus, for any network topology} system~\eqref{linsys} {can not be stable or unstable but only marginally stable}. Furthermore we point out that since no sensing/measurement {is} involved, no noise is generated from the application of the local interaction rule.

In this paper, we implemented the approach in \cite{HybridAdaptiveFreq:11} to estimate the frequencies at which the signal oscillates. Furthermore we performed a spectral analysis by means of the Discrete Fourier Transform (DFT). % Details concerning the DFT can be found in \cite{Oppenheim:1989}.

\begin{figure}%[!t]
\begin{center}
\includegraphics[width=0.45\textwidth]{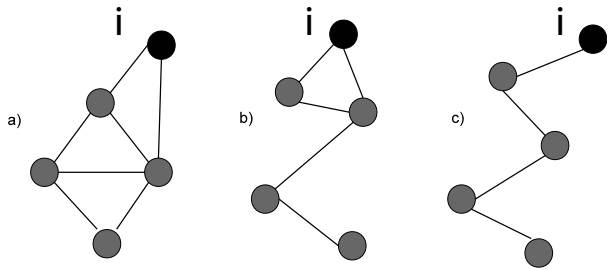}
\caption{Topology variation with respect to time for a network composed of $5$ agents.  }
\label{fig:spectrogram:1a}
\end{center}
\end{figure}
%
%\hspace{3mm}
%\begin{minipage}[b]{0.48\textwidth}
%\includegraphics[width=1\textwidth]{dccomponent.eps}
%\caption{Evolution of the average value of the agents and the estimated average value from the DC component of the FFT. }
%\label{fig:spectrogram:1b}
%\end{minipage}
%%\end{figure*}
%\begin{figure*}[!ht]
\begin{figure}
\includegraphics[width=0.45\textwidth]{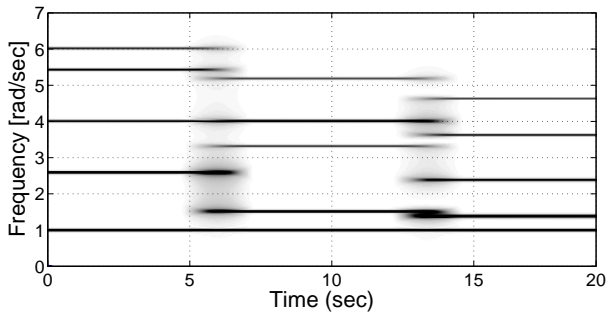}
\caption{Spectrogram of the time varying topology shown in  Fig.~\ref{fig:spectrogram:1a}  computed by the $i$th agent with respect to the output associated to its state variable $x_i(t)$. }
\label{fig:spectrogram:2}
\end{figure}
%\vspace{5mm}
%
%\begin{minipage}[b]{1\textwidth}
%\includegraphics[width=0.32\textwidth]{figs/spA.eps}\includegraphics[width=0.32\textwidth]{figs/spB.eps}\includegraphics[width=0.32\textwidth]{figs/spC.eps}
%\caption{Spectrum of the time varying topology shown in  Fig.~\ref{fig:spectrogram:1a}  computed by the $i$th agent at time $t=4$, $t=12$, $t=20$.}
%\end{minipage}
\begin{figure}
\begin{center}
\includegraphics[width=0.45\textwidth]{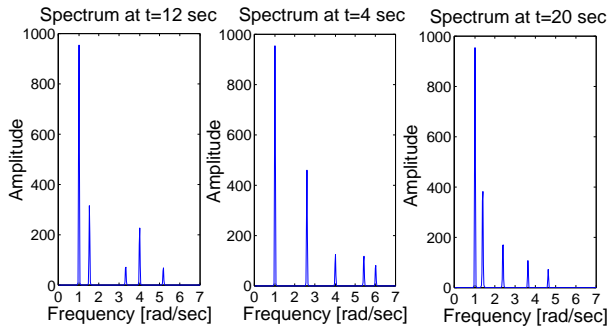}
\caption{Spectrum of the time varying topology shown in  Fig.~\ref{fig:spectrogram:1a}  computed by the $i$th agent at time $t=4$, $t=12$, $t=20$.}
\end{center}
\end{figure}

\begin{figure}[t]
\begin{center}
%\begin{minpage}[b]{1\textwidth}
\includegraphics[width=0.35\textwidth]{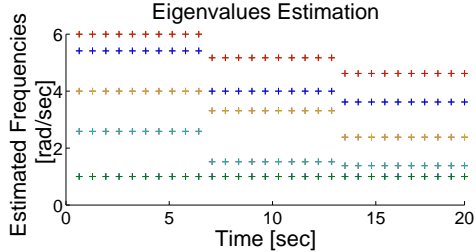}
\caption{Eigenvalues estimated every $T=1$ units of time by the approximate frequency estimation method in Subsection~\ref{methods}. Note that the frequencies are shifted by $1$ with respect to the eigenvalues of $\mathcal{L}$ in Table~\ref{table1}.} \label{figcarnevale}
\end{center}
\end{figure}

\begin{table}
\begin{scriptsize}
\begin{tabular}{|c|c|c|c|c|c|c|} \cline{1-7}
& \multicolumn{2}{|c|} {$t=5$ }& \multicolumn{2}{|c|} {$t=15$ } &\multicolumn{2}{|c|} {$t=25$ } \\ \hline
  & $\lambda_{\mathcal L}$  &  $\hat{\lambda}_{\mathcal L}$ & $\lambda_{\mathcal L}$ & $\hat{\lambda}_{\mathcal L}$ & $\lambda_{\mathcal L}$ & $\hat{\lambda}_{\mathcal L}$ \\ \hline
$\lambda_1$  & $0$ & $0$ & $0$ & $0$ & $0$ & $0$ \\ \hline
$\lambda_2$  & $1.5857$ & $1.5857$ & $0.5188$ & $0.5188$ & $0.3819$ &$0.3819$ \\ \hline
$\lambda_3$  & $3.0000$ & $2.9998$ & $2.3111$ & $2.3110$ & $1.3819$ &$1.3819$ \\ \hline
$\lambda_4$ &  $4.4142$ & $4.4138$ & $3.0000$ & $2.9998$ & $2.6180$ & $2.6179$\\ \hline
$\lambda_5$  & $5.0000$ & $4.9987$ & $4.1700$ & $5.1696$ & $3.6180$ & $3.6177$ \\ \hline
\end{tabular}
\caption{Comparison between the actual spectrum of the Laplacian, denoted by $\lambda_{\mathcal L}$, and the estimated spectrum, denoted by $\hat{\lambda}_{\mathcal L}$  of the time varying topology shown in Fig.~\ref{fig:spectrogram:1a}. }  \label{table1}
\end{scriptsize}
\end{table}
%\end{minipage}
%

\subsection{Approximate Frequencies Estimation Method} \label{methods}

The problem of estimating the frequencies of a signal {such as}
\begin{equation} \label{eq:sin}
y(t)=\sum_{i=1}^n A_isin(\omega_i t+\phi_i)
\end{equation}
has been extensively studied in control theory  via off-line methods based on Fourier analysis tools %\cite{Marple:1981}
and on-line methods \cite{Damm:1999,Loukianov:2002}. %This problem is important as it arises in a number of practical applications, e.g. sinusoidal disturbance rejection %\cite{Zhengtao:2008}.
In this paper, the approximate frequency estimation algorithm in \cite{HybridAdaptiveFreq:11} has been implemented. It allows to estimate  the frequencies of a signal of the form in eq.\eqref{eq:sin} by assuming an upper bound \texttt{nMax} to the number of existing frequencies {to be available}. The input of the algorithm is the sampling time $T_s$, an upper bound \texttt{nMax} of the number of expected frequencies, and a measure of the approximation error $S_e$. Furthermore, the length of the time window considered must be greater than the largest period of the sinusoid with the smallest frequency.

 The output of the algorithm is the number \texttt{n} of estimated frequencies and their values $\{1 + \lambda_\mathcal{L}\}$ and a flag \texttt{f} the value of which is \texttt{true} if the  percentage error when reconstructing the signal with the estimated coefficients fits the threshold $S_e$, \texttt{false} otherwise. Note that, a great advantage of this algorithm is that the estimation can be worked out in finite time. If  an observer with asymptotic convergence is required, the output of the algorithm described above could be used as input for the algorithm proposed in \cite{cd_alcosp:10}.
\subsection{Simulations with switching topology}
In order to corroborate the mathematical results, simulations have been carried out by exploiting the 4th Order Runge-Kutta Method (RK4)  to simulate the system~\eqref{linsys}. {Regarding the signal processing,} {let us recall} {that this can be always carried out locally by each agent in spite of the particular technique adopted.}

 In the simulation, a network of agents {the topology of which} changes over time is considered. In detail, Fig.~\ref{fig:spectrogram:1a}-a) depicts the network topology at the time interval  $t \in [0,6.4)$, Fig.~\ref{fig:spectrogram:1a}-b)  describes the network topology at the time interval $t \in [6.4, 12.9)$ and Fig.~\ref{fig:spectrogram:1a}-c)  describes the network topology at the time interval  $t \in [12.9, 20]$.  Each agent is running the interaction rule described in eq.~\eqref{updaterule}.

Fig.~\ref{fig:spectrogram:2} shows the spectrogram of the time varying topology computed by  the agent $i$ with respect to its state variable $x_i(t)$. The spectrogram was computed by this agent with $f_s=\dfrac{100}{2\pi}$.
The x and y axes of the spectrogram represent respectively the time step and angular frequency, while the color of the spectral line describes the amplitude of the frequency peaks, i.e., white means zero amplitude while black means an amplitude greater than $0.1$. %Note that the amplitude of the peaks in this simulations may reach $1000$.

Fig.~3 %~\ref{fig:spectrogram:3}
shows a section of the spectrogram at different time steps, namely $t=\{4, \, 12, \, 20\}$, representing the spectrum of the three network topologies taken into account. %In particular, in Table \ref{table1} are shown the eigenvalues of the Laplacian of the network topology given in Fig.~\ref{fig:spectrogram:1a}-a,b,c and the corresponding %estimated eigenvalues with the proposed algorithm. Fig.~\ref{fig:spectrogram:1b} depicts the evolution of the average value of the agents (solid line) and the estimated %average value (dashed line) from the DC component of the FFT. It can be noticed how the noise added to the system significantly affects the average value of the agents.

 To this example, we also applied the method in Section~\ref{methods} to estimate the frequency of the sinusoids and thus the eigenvalues of the Laplacian. The comparison between the eigenvalues of the Laplacian matrix of the time-varying network topology in Fig.~\ref{fig:spectrogram:1a} and the estimated eigenvalues in Fig.~\ref{figcarnevale} using the approximate frequency estimation method in Subsection~\ref{methods} is shown in Table~\ref{table1}. This method {was} implemented choosing {as sampling frequency} $f_s=\dfrac{100}{2\pi}$ that is twenty times the maximum expected frequency in the signal. This frequency corresponds to the largest eigenvalue of the Laplacian matrix plus  one. The length of the time window used to computed each estimation is $T=1$ sec, which is the period of the slowest sinusoid.
{
\subsection{Computational Cost}

To study the computational cost of Algorithm~\ref{spectrumest} we adopt the metrics proposed in \cite{Kempe200870}, i.e., we count the number of communication rounds required to obtain an estimation with a certain accuracy.  In this view, two important aspects must be considered: (i) the proposed algorithm consists in a local interaction rule that is supposed to be applied continuously; (ii) the required signal processing is carried out locally by an agent and several techniques can be adopted. Therefore, the computational cost analysis consists in the study of the cost of simulating the local interaction rule and the cost of the signal processing. Since the study of the computational cost for the signal processing required to estimate a discrete number of frequencies contained in a signal is not the scope of this paper we focus our attention to the number of communication rounds required for the discrete time simulation of system~\eqref{linsys} by the agents to collect enough data for the consecutive signal processing. In particular, the simulation time needed to collect  a sufficient amount of data must be greater than the largest period of the sinusoid with the smallest frequency, $T_{\min}$. By considering an accurate numerical simulation method such as the fourth order Runge-Kutta method, $4$ messages have to be exchanged between each agent and any of its neighbors to compute a sample of the state trajectory. It follows that for each agent the rounds of communication required to collect a sufficient amount of data  can be bounded by  $4\cdot \Delta_{\max} \cdot T_{\min} \cdot f_s$, with $\Delta_{\max}$ the maximum degree in the network and $f_s$ the chosen sampling frequency that has do be at least greater than twice the largest eigenvalue of the Laplacian, thus greater than $2\Delta_{max}$, to avoid aliasing issues.

}
\section{Conclusions} \label{sec7}
In this paper a decentralized algorithm to estimate the Laplacian spectrum {of an undirected graph} has been proposed. Each agent interacts with its neighbors so that its state oscillates at the frequencies corresponding to the eigenvalues of the Laplacian matrix that encodes the network topology. {Therefore, the problem of estimating the eigenvalues has been reduced to a simple and widely studied problem of signal processing which involves the estimation of the discrete number of frequencies at which the generated signal is oscillating. A theoretical analysis of the proposed technique along with numerical simulations has been provided.}

\bibliography{biblio}

\appendix

\section{Proof of Lemma~\ref{lemma:eigenvalues}} \label{appendix:teoeigenvectors}
By definition, the eigenvalues of $\mathcal{ A}$ are the
solutions of
$$
det( \mathcal{ A}-\lambda  I) = det\left(\left[\begin{array} {cc}
-\lambda I & I+\mathcal{ L}\\
-I-\mathcal{ L} & -\lambda I
\end{array} \right]\right)=0.$$
Since $\mathcal{ A}$ is a block matrix whose blocks commute \cite{Silvester:2000},
then
\begin{equation}\label{eig1}
det( \mathcal{ A} - \lambda I) = det\left(\lambda^2 I -\left(I+ \mathcal{L}\right)^2\right).
\end{equation}
Now, denote $\lambda_{\mathcal L}$ the generic eigenvalue of the matrix Laplacian, it is $det(\mathcal{L}-\lambda_{\mathcal L}I)=0.$
By adding and subtracting the identity matrix %we obtain
%$$det(I + \mathcal{L} -\left(1 + \lambda_{\mathcal L} \right)I)=0.$$
and exploiting the fact that the eigenvalues of the square of a matrix are squared, we obtain
$$det(\left(I + \mathcal{L} \right)^2-\left(1 + \lambda_{\mathcal L} \right)^2I)=0.$$
Thus, by \eqref{eig1}, it is
$$\lambda^2=-\left(1 + \lambda_{\mathcal L} \right)^2, \quad \Rightarrow \quad \lambda=\pm j \left(1 + \lambda_{\mathcal L} \right).$$
Therefore, to each eigenvalue of the Laplacian matrix ${\mathcal L}$ it corresponds two imaginary and conjugate eigenvalues of matrix ${\mathcal A}$ denoted by $\lambda_{\mathcal A}$ and $\bar{\lambda_{\mathcal A}}$, so that
$$\lambda_{\mathcal A}=j\left(1 + \lambda_{\mathcal L} \right),\quad \text{and} \quad \bar{\lambda}_{\mathcal A}=-j\left(1 + \lambda_{\mathcal L}\right),$$
%
%Hence,
%%$$\begin{array} {ll}
%%det \left( \mathcal{ L}^2 +  \lambda^2 I  \right) &= det \Big(\left( \mathcal{ L} +  j \lambda I  \right)\left(  \mathcal{ L}  -j\lambda I\right)\Big)\\ [6pt]
%%&= det\left( \mathcal{ L} +  j \lambda I  \right)det\left( \mathcal{ L} -j  \lambda I  \right)  = 0.
%%\end{array}$$
%$$
%det \left( \mathcal{ L}^2 +  \lambda^2 I  \right)
%= det\left( \mathcal{ L} +  j \lambda I  \right)det\left( \mathcal{ L} -j  \lambda I  \right)  = 0.
%$$
%
%Let us now recall that $\lambda_\mathcal{ L}$ is the generic (real and positive) eigenvalue of the Laplacian matrix $L$, that is
%\mbox{$det ( \mathcal{L} - \lambda_{\mathcal{L}} I) = 0$},
%
%which are all real and positive. Therefore, by substituting $\lambda_\mathcal{ L} = - j \left(\lambda+1\right)$ within the term $det\left( \mathcal{ L} +  j \left(\lambda+1\right)I  \right)$  and
%$\lambda_\mathcal{ L} =  j \left(\lambda+1\right)$ within the term $det\left( \mathcal{ L} -  j \left(\lambda+1\right) I  \right)$, we obtain:
%$\lambda_\mathcal{ A}= j\left(\lambda_{\mathcal{L}}+1\right), \bar \lambda_\mathcal{ A}=-j\left(\lambda_{\mathcal{L}}+1\right)$
thus proving the first statement. Now, by definition, the eigenvectors of $\mathcal{A}$ relative to $\lambda_{\mathcal{A}}$ are solutions of
$$
\left[\begin{array} {cc}
\mathbf{0}_{n \times n} & I+\mathcal{ L}\\
-I-\mathcal{ L} & \mathbf{0}_{n \times n}
\end{array} \right]
\cdot \left [
\begin{array}{l}
{ v'} \\
{ v''}
\end{array}
\right] = \lambda_{\mathcal{ A}} \left [
\begin{array}{l}
{ v'} \\
{ v''}
\end{array}
\right]
$$
%$$
%\left \{
%\begin{array}{r}
%I+\mathcal{ L} \, { v''} =j\, \lambda_{\mathcal{ L}}+1 \, { v'} \\
%-I-\mathcal{ L} \, { v'}=j\, \lambda_{\mathcal{ L}}+1\,  { v''}
%\end{array}
%\right.
%$$
%$$
%\left \{
%\begin{array}{r}
%\left(I+\mathcal{ L} \right)\, { v''} =j\, \left(1 + \lambda_{\mathcal{ L}} \right) \, { v'} \\
%\left(-I-\mathcal{ L} \right)\, { v'}=j\, \left(1 + \lambda_{\mathcal{ L}} \right)\,  { v''}
%\end{array}
%\right.
%$$
%
for which a possible solution is
$v_{\mathcal{A}}=[v_{\lambda_{\mathcal{ L}}}^{T}\quad j v_{\lambda_{\mathcal{ L}}}^{T} ]^{T}$.
 The same argument holds for the conjugate eigenvalue $ \bar \lambda_{\mathcal{ A}} = - j\,
\left(1+ \lambda_{\mathcal{ L}} \right)$ for which a possible solution is \mbox{$\bar{v}_{\mathcal{A}}=[v_{\lambda_{\mathcal{ L}}}^{T} \; -j
v_{\lambda_{\mathcal{ L}}}^{T} ]^{T}$}.

\section{Proof of Theorem \ref{teo5}} \label{appendix:teo5}

%The state trajectory of $x_i(t)$ is a linear combination of
%the system modes.
When referring to the eigenvalues and
eigenvectors of $\mathcal{L}$, $\lambda_{\mathcal{ L},j}$ and
$v_{\lambda_{\mathcal{L},j}}$ for $j=1,\ldots,n$, we drop the
subscripts $\mathcal{L}$ and $\lambda_\mathcal{L}$, respectively,
and refer to them as $\lambda_j$ and $v_{j}$ for
$j=1,\ldots,n$.
 %Since $\mathcal{ A}$ is skew symmetric, and any skew
%symmetric matrix is a normal matrix, i.e. $\cal{A}^{\ast}
%\cal{A} = \cal{A} \cal{A}^{\ast}$, thanks to the Spectral
%Theorem it is always diagonalizable through a unitary
%matrix\footnote{A unitary matrix $U$ is a complex matrix such that
%$U^*U=UU^*=I$, where $U^*$ is the complex conjugate of $U$. }. Thus
%all the eigenvalues have geometric multiplicity equal to their
%algebraic multiplicity (or equivalently, unitary index).
 By Lemma~\ref{lemma:eigenvalues} to each Laplacian eigenvalue
$\lambda_j$ it corresponds a couple of pure imaginary eigenvalues of
$\mathcal{ A}$ equal to
$$\lambda_{\mathcal{ A}},\bar \lambda_{\mathcal{ A}}= \pm
j \left(1 + \lambda_j \right).$$
Therefore, the state trajectory $x_{i}(t)$ of each agent
 is
%$$ x_i(t)=\sum_{j=1}^{m} a_{j,i}\, \sin(\left(1 + \lambda_j \right) t + \phi_j),$$
%
a linear combination of sinusoids whose amplitudes and phase shifts are function
of the initial conditions and of the graph topology.

 Now, we compute the coefficients of the module of the Fourier transform of $x_i(t)$.
%When referring to the eigenvalues and
%eigenvectors of $\mathcal{ L}$, $\lambda_{\mathcal{ L}_i}$ and
%$v_{\lambda_{\mathcal{ L}_i}}$ for $i=1,\ldots,n$, we drop the
%subscripts $\mathcal{ L}$ and $\lambda_\mathcal{ L}$, respectively,
%and refer to them as $\lambda_i$ and $v_{ i}$ for
%$i=1,\ldots,n$.
Since $\mathcal{A}$ is skew symmetric, and any skew
symmetric matrix is a normal matrix, i.e. $\mathcal{A}^{\ast}
\mathcal{A} = \mathcal{A} \mathcal{A}^{\ast}$, thanks to the Spectral
Theorem it is always diagonalizable through a unitary
matrix\footnote{A unitary matrix $U$ is a complex matrix such that
$U^*U=UU^*=I$, where $U^*$ is the complex conjugate of $U$. } and
all the eigenvalues have geometric multiplicity equal to their
algebraic multiplicity (or equivalently, unitary index).
Thus $\mathcal{ A}$ can be decomposed as $\mathcal{A} =VDV^{\ast}$,
where $D$ is a diagonal matrix whose elements are arranged as
\mbox{$ D= \text{diag} \{ j\lambda_{1}, \;\; \ldots, \;\; j\lambda_{n}, \;\; -j\lambda_{1}, \;\; \ldots, \;\; -j\lambda_{n} \}
$},
and $V$ is a complex matrix whose columns are the eigenvectors of
$\mathcal{ A}$. Furthermore, applying Lemma \ref{lemma:eigenvalues},
matrix $V$ is rearranged to match the disposition of the eigenvalues
of $D$ as follows
$$V= \left[\begin{array}{cccccccc}
v_{1} & v_{2}& \ldots & v_{n} & v_{1} & v_{2}& \ldots & v_{n}\\
jv_{1} & jv_{2} & \ldots  &j v_{n} & -jv_{1} & -jv_{2}& \ldots & -jv_{n}\\
\end{array}\right].$$
In the following it is assumed that $v_{j}$, $j=1,\ldots,n$, are
normalized eigenvectors such that $\|v_{j}\|=1$ and $VV^\ast=I$.
%To keep this notation we need to normalize the eigenvectors of $\mathcal{ A}$
%such that $VV^\ast=I$,
Thus
$$\left\|\alpha \Big[
v_{j}^{T} \quad
jv_{j}^{T} \Big ]^{T} \right \|=1, \quad \alpha\in \rea.$$
%$$\left\|\alpha \left[\begin{array}{c}
%v_{i}\\
%jv_{i}  \end{array}\right]\right\|=1.$$
By simple manipulations we
find $\alpha= \frac{1}{\sqrt{2}}$.
The state trajectories of the system are captured by the matrix exponential which in our case takes the form $e^{\mathcal{ A}t} = Ve^{Dt}V^\ast$. \\
%
% It follows that the state trajectory of $x(t)$ and $z(t)$ have the following form
%$$
%\left[\begin{array} {c}
%x(t)\\
%z(t)
%\end{array}\right]=\\
%\frac{1}{2}Ve^{Dt}V^\ast\left[\begin{array}{c}
%x(0)\\
%z(0)
%\end{array}\right],
%$$
%By some manipulations we find
 It follows that the state trajectory of $x_i(t)$ and $z_i(t)$ have the following form
$$
\begin{aligned}
x_i(t) &= \sum_{j=1}^{n} \left[v_{j}(i)\left(\cos(\left(1 + \lambda_{j} \right)t)v_{j}^Tx(0)\right.\right. \\
        &+\left.\left.\sin(\left(1 + \lambda_{j} \right)t)v_{j}^Tz(0)\right) \right], \\[5pt]
z_i(t)&= \sum_{j=1}^{n} \left[v_{j}(i)\left(-\sin(\left(1 + \lambda_{j} \right)t)v_{j}^Tx(0)\right.\right. \\
       &+\left.\left.\cos(\left(1 + \lambda_{j} \right)t)v_{j}^Tz(0)\right)\right].
\end{aligned}
$$
{Thus, according to the notation of Theorem~\ref{teo5}, by simple manipulations we can obtain the expression for the coefficients $a_{j,i}$ and $b_{j,i}$ associated to the eigenvalue $\lambda_{j}$ for the $i$-th agent stated in eq.~\eqref{coefficients} and eq.~\eqref{coefficients2}, proving the theorem.}
%%
% -- For $\lambda_{ 1} = 0$  and $\nu_{ 1} = 1$
%$$
%\begin{aligned}
%a_{ 1,i} &= v_{ 1}(i) \, v_{ 1}^Tx(0){
%=}\dfrac{1}{ \sqrt{n}}\dfrac{\textbf{1}_{n}^Tx(0)}{\sqrt{
%n}}=\dfrac{\textbf{1}_{n}^Tx(0)}{n} \\[5pt]
%%|\mathcal{F}\left[z_i(t)\right]|
%b_{ 1,i} &=   v_{ 1}(i) \, v_{ 1}^Tz(0) =
%\dfrac{1}{\sqrt{n}}\dfrac{\textbf{1}_{n}^Tz(0)}{\sqrt{n}}=\dfrac{\textbf{1}_{n}^Tz(0)}{n}.
%\end{aligned}
%$$
%%
% where $v_1=\frac{1}{\sqrt{n}}\textbf{1}_{n}$ is the unitary norm
%eigenvector associated to $\lambda_1$. \\
%%
% -- For $\lambda_j  \geq 0$ and $\nu_{j} \geq 1$
%%
%%$$
%%\begin{array}{ll}
%%%|\mathcal{F}\left[z_i(t)\right]|
%%%\dfrac{b_{j,i}}{2} &=& v_{\lambda_{j}}(i) \, \sqrt{(v_{\lambda_{j}}^Tx(0))^2 + (v_{\lambda_{j}}^Tz(0))^2}.
%%a_{j,i}=b_{j,i}=& \sqrt{ \left [ \sum_{k=1}^{\nu_{j}} \Big (v^{(k)}_{j}(i)    {v^{(k)}_{j}}^Tx(0) \Big ) \right ]^{2}} +\\
%%\vspace{-0.3cm} \\
%%& \overline{+\left [\sum_{k=1}^{\nu_{j}} \Big ( v^{(k)}_{j}(i) {v^{(k)}_{j}}^Tz(0) \Big) \right ]^{2} }.
%%\end{array}
%%$$
%$$
%%|\mathcal{F}\left[z_i(t)\right]|
%%\dfrac{b_{j,i}}{2} &=& v_{\lambda_{j}}(i) \, \sqrt{(v_{\lambda_{j}}^Tx(0))^2 + (v_{\lambda_{j}}^Tz(0))^2}.
%{a_{j,i}=b_{j,i}=
%\sqrt{
%\left [ \sum_{k=1}^{\nu_{j}} \Big (v^{(k)}_{j}(i)    {v^{(k)}_{j}}^Tx(0) \Big ) \right ]^{2} + \atop \left [\sum_{k=1}^{\nu_{j}} \Big ( v^{(k)}_{j}(i) {v^{(k)}_{j}}^Tz(0) \Big) \right ]^{2}
%}.}
%$$

\section{Proof of Theorem~\ref{teo:obs}} \label{appendix:teo:obs}
{Let $\mathcal{O}_{\mathcal{A}}=\mathcal{O}(\mathcal{A}, \hat{C})$, $\mathcal{O}_{I+\mathcal{L}}=\mathcal{O}(I+\mathcal{L}, C)$ and {$\mathcal{O}_{\mathcal{L}}=\mathcal{O}(\mathcal{L}, C)$} be the observability matrices of the corresponding matrices.
It can be shown by  row permutation that:
$$ \begin{array} {l}
\text{Rank} \left( \mathcal{O}_{\mathcal{A}} \right)= \\
\text{Rank} \left( \left[ \begin{array}{cc}
                    \mathcal{O}_{I+\mathcal{L}} & \textbf{0}_{n \times n} \\
                    \mathcal{O}_{I+\mathcal{L}}(I+\mathcal{L})^n & \textbf{0}_{n \times n} \\
                     \textbf{0}_{n \times n} & \mathcal{O}_{I+\mathcal{L}} \\
                    \textbf{0}_{n \times n} &\mathcal{O}_{I+\mathcal{L}}(I+\mathcal{L})^n \\
                  \end{array} \right] \right)
                  \end{array}$$
hence it also holds that
$$\text{Rank} \left( \mathcal{O}_{\mathcal{A}} \right)= \\
\text{Rank} \left( \left[ \begin{array}{cc}
                    \mathcal{O}_{I+\mathcal{L}} & \textbf{0}_{n \times n} \\
                     \textbf{0}_{n \times n} & \mathcal{O}_{I+\mathcal{L}} \\
                  \end{array} \right] \right)$$
from which it follows that $Rank\left(\mathcal{O}_{\mathcal{A}}\right)=2 Rank\left(\mathcal{O}_{I+\mathcal{L}}\right)$.\\

 Finally, by noticing that the eigenvalues of the matrices $\mathcal{O}_{I+\mathcal{L}}$ and $\mathcal{O}_{\mathcal{L}}$ are related as $\lambda_{I+\mathcal{L}} =  1 + \lambda_{\mathcal{L}}$  and  share the same eigenvectors, from the  PBH observability test it follows that
$$Rank \left(\mathcal{O}_{\mathcal{L}}\right)= Rank\left(\mathcal{O}_{I+\mathcal{L}}\right),$$
thus
$$Rank\left(\mathcal{O}_{\mathcal{A}}\right)=2 Rank\left(\mathcal{O}_{\mathcal{L}}\right).$$}

\end{document}